\documentclass[sigconf]{acmart}
\acmConference[EASE 2025]{The 29th International Conference on Evaluation and Assessment in Software Engineering}{17–20 June, 2025}{Istanbul, Türkiye}

\AtBeginDocument{%
  }

\setcopyright{acmlicensed}
\copyrightyear{2025}
\acmYear{2025}
\acmDOI{XXXXXXX.XXXXXXX}

\usepackage{colortbl}
\usepackage{threeparttable}
\usepackage{graphicx}
\usepackage{balance}
\usepackage{titlesec}
\titlespacing*{\paragraph}{0pt}{0.9ex plus .2ex minus .2ex}{0.5ex plus .2ex}

\begin{document}

\title{Exploring turnover, retention and growth in an OSS Ecosystem}


\author{Tien Rahayu Tulili, Ayushi Rastogi, Andrea Capiluppi}
\affiliation{%
  \institution{Bernoulli Institute for Mathematics, Computer Science, and Artificial Intelligence} 
  \city{Groningen}
  \state{Groningen}
  \country{The Netherlands}
  }
\email{{t.r.tulili, a.rastogi, a.capiluppi}@rug.nl}

\acmArticleType{Research}
\acmCodeLink{https://github.com/borisveytsman/acmart}
\acmDataLink{htps://zenodo.org/link}

\begin{abstract}
The Gentoo ecosystem has evolved significantly over 23 years, highlighting the critical impact of developer sentiment on workforce dynamics such as turnover, retention, and growth. While prior research has explored sentiment at the project level, sentiment-driven dynamics at the component level remain underexplored, particularly in their implications for software stability.

This study investigates the interplay between developer sentiment and workforce dynamics in Gentoo. The primary objectives are to (1) compare workforce metrics (turnover, retention, and growth rates) between sentiment-positive (SP) and sentiment-negative (SN) components, (2) examine temporal trends across three time phases, and (3) analyze the impact of these dynamics on software stability.

A mixed-method approach was employed, integrating sentiment analysis of mailing lists and commit histories using the SentiStrength-SE tool. Workforce metrics were statistically analyzed using Pearson Correlation Matrix and Mann-Whitney U tests. The analysis focused on the most SP and SN components in the ecosystem.

SN components exhibited higher retention rates but slower growth and turnover compared to SP components, which showed dynamic contributor behavior but reduced long-term stability. Temporal analysis revealed significant variations in workforce dynamics over three phases, with developer retention correlating positively with modifications in both sentiment groups.

Tailored strategies are necessary for managing sentiment-driven dynamics in OSS projects. Improving \textit{adaptability} in SN components, and \textit{continuity} in SP components, could improve project sustainability and innovation. This study contributes to a nuanced understanding of sentiment's role in workforce behavior and software stability within OSS ecosystems.
\end{abstract}


\maketitle
\section{Introduction}

An open-source software (OSS) ecosystem is a collaborative environment where developers, users, and organizations contribute to, maintain, and benefit from open-source software projects, fostering shared resources and innovation. Given the OSS ecosystem's reliance on voluntary contributors, its dynamics are shaped by developer engagement, retention, and turnover.

Gentoo Linux exemplifies an OSS ecosystem that has faced challenges related to retention and turnover: Garcia et al.~\cite{Garcia2013Role} conducted a study exploring factors that drive developers to leave, focusing on sentiments expressed in Bugzilla reports and mailing lists. Using the SentiStrength tool, they identified emotional intensity as a potential predictor of developer turnover. Similarly, unexpected departures of core developers have been shown to significantly impact organizational structures and the performance of teams, highlighting the need for stability and support~\cite{zanetti2013rise}.

Workforce dynamics, such as turnover and retention, can pose risks to the sustainability of OSS projects. The survival of an OSS community is intricately linked to its projects, which often consist of interdependent components. High contributor turnover may pose a risk to the integrity and viability of these essential components, increasing project costs and delays.

Investigating sentiment at the component level introduces unique challenges compared to project-level analyses. Components often function as modular, reusable units developed and maintained by distinct teams or contributors, adding complexities such as varying commitment levels. Adopting open-source components also entails challenges like compatibility, security, and ongoing maintenance~\cite{  TheScaleOfOpen2025}, all of which can influence contributor sentiment and affect individual components.

To explore collaboration dynamics at the component level, our work focuses on the Gentoo Linux ecosystem. Established in 1999, Gentoo is a large OSS project organized into modular categories, providing a natural framework for studying sentiment-driven workforce metrics. While prior studies have examined developer sentiment in GitHub commit logs~\cite{Sinha2016Analyzing} and email threads~\cite{Ferreira2019Longitudinal}, they primarily focus on broader project-level trends. For instance, Sinha et al.~\cite{Sinha2016Analyzing} analyzed commit messages for emotional patterns, while Ferreira et al.~\cite{Ferreira2019Longitudinal} studied shorter-term email sentiments. These studies often overlook finer-grained insights at the component level, leaving workforce metrics like turnover, retention, and growth rates underexplored.


Behavioral studies in OSS further highlight the interplay between developer activity and project stability. For example, Robinson et al.~\cite{Robinson2016Developer} analyzed sentiment-related behaviors across projects, while Zhou and Mockus~\cite{Zhou2012What} identified drivers of sustained contributions. However, these studies neglect sentiment-driven dynamics at a granular level, such as specific components within a project. The implications of turnover and retention on component changes, and the differences between sentiment-positive (SP) and sentiment-negative (SN) categories, remain unexplored.


This study aims to fill these gaps by analyzing developer dynamics in the Gentoo ecosystem over a 23-year period (2000–2023). We integrate two datasets (mailing list communications and commit histories) to explore three workforce metrics: turnover, retention and growth rates. Specifically, we aim to analyze the following:
\begin{itemize}
    \item Differences in turnover, retention and growth rates between SP and SN components.
    \item Temporal trends in workforce dynamics across distinct time periods.
    \item The impact of workforce metrics (e.g., retention, turnover, and growth rates) on the stability of sentiment-affected components, specifically to the modifications of the components.
\end{itemize}

Unlike prior research that relies on single datasets or project-level analyses, this study adopts a deeper, component-level perspective. This approach provides nuanced insights into how developer sentiment influences workforce dynamics and software stability, addressing critical gaps in the literature.

This paper is organized as follows: Section~\ref{sec:_related_work} reviews work on sentiment analysis and developer dynamics in OSS. Section~\ref{sec:_methodology} outlines definitions, Goal-Question-Metrics, and data methods. Section~\ref{sec:_results} presents findings on workforce metrics and their relationship with sentiment-affected components. Section~\ref{sec:_discussion_and_implications} discusses implications for OSS management, while Section~\ref{sec:_threats_to_validity} addresses validity threats. Finally, Section~\ref{sec:_conclusions} concludes the paper and suggests directions for future research.

\vspace{-0.5em}

\section{Related Work}
\label{sec:_related_work}


\paragraph{Developers, turnover, retention, and growth}
Long-term engagement factors were explored by Zhou and Mockus \cite{Zhou2012What}, who identified drivers of sustained contributions in open-source communities. The factors in their study included social aspects (e.g. helping others, teamwork, and reputation), learning, and intellectual stimulation. Furthermore, Chaves et al. \cite{chaves2022autonomy} reviewed autonomy and turnover dynamics. In this literature review work, they pointed out the autonomy factors linked to the turnover developers: employees' wishes, communication contributor's opinion, individual factors, career satisfaction, job opportunities, work regime, and work routine. 

Developer turnover poses significant challenges to project continuity and knowledge retention. Hall et al. \cite{T2008impact} and Izquierdo-Cortazar et al. \cite{Daniel2009Using} highlighted the implications of turnover, suggesting the need for effective mitigation strategies. Kaur et al. \cite{Kaur2022Analysis} connected turnover patterns with sentiment fluctuations in commit logs, offering predictive insights to improve workforce management.

Previous studies have investigated retention among software developers. Triekenreich et al.~\cite{trinkenreich2024predicting}, proposed a prediction model of attrition by considering factors derived from the Job Demands-Resources (JD-R) model. They found that engagement is a stronger predictor of retention than burnout and intentions to stay. Organizational culture and learning opportunities also significantly predict engagement and burnout across all demographics. Developers with opportunities to stimulate personal learning and development will be more engaged to work~\cite{stol2022gamification}. In the model of motivation in software engineering proposed by Sharp et al.,~\cite{sharp2009models}, they reported that one of the positive outcomes of motivated software engineers is retention.

Meanwhile, Koch [12] analyzed software evolution trends at a macro level within open-source  ecosystems, specifically in terms of the growth rate. Their study explored the evolutionary behaviour with statistical models. One of their findings was larger projects with a higher number of contributions are more likely to achieve and sustain super-linear growth.

\paragraph{Developers and sentiments}
Past studies highlight the impact of developors’ emotions on their performance during software development. Wrobel~\cite{wrobel2013emotions} conducted an online survey of developers that revealed that negative emotions such as anger may impact on developers' productivity. In addition, negative emotions such as frustration and disgust may pose significant risks that should not be overlooked. These findings emphasize the importance of addressing emotional states in software development to mitigate potential risks and leverage emotions that may enhance productivity. 

To minimize subjective bias when assessing emotions in this context, researchers have explored alternative data sources, including biometric inputs~\cite{chandler2012biometric, girardi2017emotion}, keyboard interaction patterns~\cite{vizer2009automated}, and the analysis of audio, video~\cite{ liu2018speech}, and written text~\cite{ gachechiladze2017anger}. Despite all aforementioned studies, none have considered developers' sentiments at the component level, which represents the foundational elements of the project itself and serves as the collaborative space where developers work together. 


Guzman et al. \cite{Guzman2014Sentiment} studied GitHub commit logs to explore emotional patterns in software development. They considered factors such as programming language and commit timing, with emotions measured using SentiStrength. Their analysis focused on weekly and project-based trends from commit messages. Sinha et al. \cite{Sinha2016Analyzing} expanded on this by analyzing over 2 million commits and linking sentiments to file changes within projects. In our study, we broaden the dataset by including commit logs and technical discussions from mailing lists over 20 years, and we focus on sentiment at a more detailed level, examining specific components of the projects. 

Ferreira et al.~\cite{Ferreira2019Sentiment} analyzed over 15k email messages from Linux subsystem maintainers and focused on sentiment changes around significant events like the Linux community's temporary break. They used Senti4SD for sentiment analysis and examined trends over different time periods—monthly, weekly, and daily. Similarly, our work utilized technical-related messages to conduct sentiment analysis; however, our data observation ranged over 23 years and explored the sentiments and the workforce metrics at the component level based on three different phases.

Expanding the scope, Calefato et al. \cite{Calefato2018On} introduced Senti4SD and analyzed Stack Overflow posts to link personality traits with sentiments. Garcia et al. \cite{Garcia2013Role} demonstrated how emotional content in Bugzilla reports and mailing lists influenced contributor activity in the Gentoo Linux ecosystem. They used a Bayesian approach to predict when contributors might become inactive. Similar to our work, we conducted the sentiment analysis by extending the time period into 23 years at the component level. We explore workforce metrics, including retention, turnover, and growth rates to study the potential relationship with the commit activity.

Behavioral dimensions of developers have also been studied in the context of project outcomes. Robinson et al. \cite{Robinson2016Developer} mined repository data from 124 projects to explore sentiment-related developer behaviors. Majumder et al. \cite{Majumder2022Revisiting} provided a process- and product-based perspective on developer impact using metrics-based analysis. Similarly, we studied sentiment-related developer behavior, except we dived more into the component level. We focused on the workforce metrics (e.g. turnover, retention, and growth rates). 

\paragraph{Developers and components}
Wu et al.~\cite{wu2023social} studied social and technical dependency networks in open-source software (OSS) communities. They analyzed network metrics like degree centrality and betweenness centrality to assess their impact on project success. The findings revealed nonlinear relationships, showing that more connections do not always lead to greater OSS success. However, this study did not consider sentiment as one of the variables to be measured in the social dependency networks.

Palyart et al. \cite{palyart2017study} explored social interactions between developers of projects using a component and the developers of that component. Their study revealed that as a component is used more, interactions between these developers become less frequent and shorter in duration. However, while they focused on project-level interactions, our study examines the component level and incorporates sentiments.

Nucci et al.~\cite{di2017developer} conducted a study that incorporated both developers and components in a bug prediction model, finding that analyzing the distribution of changes made by developers improved predictions. Previous research has also highlighted factors related to developers, such as the number of developers on a component~\cite{bell2013limited}, change-proneness analysis~\cite{moser2008analysis}, and change entropy~\cite{hassan2009predicting}. These studies indicate that defects are more likely to occur at the component level, highlighting the need for further investigation. Accordingly, our study explores the sentiments of developers regarding components and includes workforce metrics like turnover, retention, and growth rates.


\section{Methodology}
\label{sec:_methodology}
We selected the Gentoo community, a large open-source ecosystem, for our study due to its previously noted instances of negative communication influencing developer behavior~\cite{Garcia2013Role}. Established in 1999, the Gentoo Linux project is still active, focusing on a free Linux-based operating system. It is committed to its core values of customization, optimization and community involvement in development. 

Gentoo is a source-based Linux distribution that allows users to build and customize their systems. It manages a variety of open-source components using Portage, its package manager, which handles fetching, compiling and installing software. As of this study, Gentoo offers 19,034 packages across 170 categories on its official website and is involved in 111 projects. Our research focused on the packages available in these categories on GitHub.

To answer our research questions, we followed several steps, starting with data collection, data preprocessing, and data analysis. We describe the description of each stage as follows:

\subsection{Definitions}
\label{sec:_definitions}
To ensure clarity and consistency, this study employs the following definitions:
\begin{itemize}
    \item \textbf{Category}: A high-level grouping of files within the codebase. For instance, the file `\textit{media-libs/id3lib/files/digest-id3lib-3.8.0\textunderscore pre2}` belongs to the `\textit{media-libs}` category. Gentoo’s official website\footnote{https://packages.gentoo.org/categories} lists 170 distinct categories as of our analysis started, each serving a specific purpose. In this study, we use the term `\textit{category}' to represent a Gentoo component.
    
    \item \textbf{Sentiment-negative (SN) categories}: Categories exhibiting predominantly negative sentiment in the communication between developers. Our study focuses on 10 ten most negative categories. The details of how we selected the categories is explained in Section~\ref{sec:_data_analysis}.
    
    \item \textbf{Sentiment-positive (SP) categories}: Categories exhibiting predominantly positive sentiment. Our study focuses on the ten most positive categories. The details of how we selected the categories is explained in Section~\ref{sec:_data_analysis}.
    
    \item Our study's observation covers the time period between 2001 and 2023. In order to help us find out the pattern to answer our research questions, we divided this period into three time phases:
        \begin{itemize}
        \item \textbf{Early phase:} The period from 2001 to 2007.
        \item \textbf{Middle phase:} The period from 2008 to 2014.
        \item \textbf{Last phase:} The period from 2015 to 2023.
    \end{itemize}
    
    \item \textbf{Time intervals:}
    \begin{itemize}
        \item \textbf{$\delta_1$:} Comparison between the early and middle phases.
        \item \textbf{$\delta_2$:} Comparison between the middle and last phases.
        \item \textbf{$\delta_3$:} Comparison between the early and last phases.
    \end{itemize}

    \item \textbf{Growth rate (GR)}: The percentage of developers who contributed to a category for the first time in a given month, calculated as:
    \vspace{-0.75em}
    \begin{quote}
    \small
        \[
    \text{GR} = \frac{\text{Total new developers joining in a month}}{\text{Total developers in all categories in a month}} \times 100;
    \]
    \end{quote}
    We adopted the concept of growth in the work of Cooper et al.,~\cite{cooper1991resource} and defined our growth rate as described above.
    \item \textbf{Retention rate (RR)}: The percentage of developers who actively contributed commits to a specific category in a given month, calculated as:
    \vspace{-0.75em}
    \begin{quote}
    \small
        \[
    \text{RR} = \frac{\text{Total developers active in a month}}{\text{Total developers in all categories in a month}} \times 100;
    \]
    \end{quote}
    We adopted the retention concept as described in the work of Shen et al.~\cite{shen2004factors} and defined our retention rate by adopting, with modification, the formula described in the work of Moscelli et al.~\cite{moscelli2025staff}.

    \item \textbf{Turnover rate (TR)}: The percentage of developers who stopped contributing to a category in a given month, calculated as:
    \vspace{-0.75em}
    \begin{quote}
    \small
        \[
    \text{TR} = \frac{\text{Total developers stopped contributing in a month}}{\text{Total developers in all categories in a month}} \times 100;
    \]
    \end{quote}
    We modified the formula of TR proposed by Arthur~\cite{arthur1994effects} and Guthrie~\cite{guthrie2001high} by adding a specific range of time.

    \item \textbf{Number of modifications (M)}: The number of changes made by developers by utilizing the following formula:
    \begin{quote}
    \small
        \[
    \text{M} = (log(N) * RR * log(C\_retained)) + (log(G) * log(C\_new)); 
    \]
    
    \end{quote}
    where N is the number of developers contributing to a category in a month; C\_retained is the average of modifications made by the developers N in a month; G is the number of new developers in a month; C\_new is the average of modifications made by new developers G in a month.
\end{itemize}

\subsection{Goal-Questions-Metrics (GQM)}
\label{sec:_gqm}
The objective of this study is to examine three key workforce metrics—turnover, retention and growth rates in an open-source ecosystem, with a specific focus on sentiment-affected components. The study aims to explore the relationship between these workforce metrics and the modifications made in the software components. Using the Goal-Questions-Metrics (GQM) approach, the research goal is defined as follows:

\begin{quote}
\textbf{To examine} turnover, retention and growth rates within sentiment-affected \textbf{categories} in the Gentoo community over the past 23 years. \textbf{To investigate} the relationship between these workforce metrics and the modifications made in the sentiment-affected components.
\end{quote}

Based on this goal, the research is structured around two primary questions. The first question (RQ1) focuses on the dynamics of turnover, retention and growth across SN and SP components, while the second question (RQ2) investigates the impact of these dynamics on component (or category) stability.

\textbf{RQ1 (a)}: Are there differences in turnover, retention and growth rates between SN and SP categories?  

\textit{Rationale -} This question seeks to determine whether turnover, retention, and growth rates differ systematically between components associated with predominantly positive or negative sentiments. Our results show that the differences between components exhibiting predominantly negative sentiments and positive sentiments statistically existed.

\textbf{Metrics:} The analysis uses metrics such as the number of new developers contributing to categories, the number of developers leaving categories and the number of active developers within categories. Sentiment classification is conducted using the Sentistrength-SE tool~\cite{islam2018sentistrength}.

\textit{Statistical testing:}  The null hypotheses to be tested are: \begin{itemize}
    \item $H_{RQ1\textbf{a}_1,0}$: “\textit{There is no difference in the turnover rate between the most positive categories and negative categories}”;
    \item $H_{RQ1\textbf{a}_2,0}$: "\textit{There is no difference in the retention rate between the most positive categories and negative categories}";
    \item $H_{RQ1\textbf{a}_3,0}$: "\textit{There is no difference in the growth rate between the most positive categories and negative categories}".
\end{itemize} 

\noindent\makebox[\columnwidth]{\rule{\columnwidth}{0.4pt}}

\textbf{\textit{RQ1 (b)}} - Are there particular time periods where retention, turnover and growth rates differ significantly between SP and SN categories? 

\textit{Rationale -} This question explores temporal trends in turnover, retention, and growth rates, aiming to identify whether differences between sentiment groups persist or vary across distinct time periods. By doing so, it highlights the evolution of workforce patterns over the early, middle and latter phases of the project. Our results show that the differences between the two groups persists across three time periods (e.g. early, middle, and last phase). 

\textit{Metrics:} Similar to RQ1 (a), metrics include the number of developers joining, leaving, or actively contributing to categories, sentiments are classified using the Sentistrength-SE tool.

\textit{Statistical testing:}  The null hypotheses to be tested are: \begin{itemize}
    \item $H_{RQ1\textbf{b}_1,0}$: “\textit{There is no difference in the turnover rate between each time period of the most SN categories and SP categories}”;
    \item $H_{RQ1\textbf{b}_2,0}$: "\textit{There is no difference in the retention rate between each time period of the most SN categories and SP categories}";
    \item $H_{RQ1\textbf{b}_3,0}$: "\textit{There is no difference in the growth rate between each time period of the most SN categories and SP categories}".
\end{itemize} 

\noindent\makebox[\columnwidth]{\rule{\columnwidth}{0.4pt}}

\textbf{\textit{RQ2}} - Do retention rate, turnover and growth rate correlate with modifications in sentiment-affected components?

\textit{Rationale -} This question investigates the correlation between turnover, retention, and growth rates and software component modifications. Understanding this correlation can help predict which components are most vulnerable to instability due to turnover or retention issues. Our results show that the growth drives an increased modification activity in the SP components; however, the growth was correlated negatively with the modifications made in the SN components. Additionally, the modifications do not correlate with turnover; however, they have strong positive correlations with retention in both groups.

\textit{Metrics:} Key metrics include the number of new developers, departing developers and number of developers who contributed on each month in each component, as well as the frequency of commits and updates to components over time to obtain the number of modifications made by developers. Sentiment-affected components are classified using the Sentistrength-SE tool.

\subsection{Data collection}
This study used two primary datasets to examine the dynamics of developer sentiment and activity within the Gentoo open-source ecosystem. These datasets include: 

\begin{enumerate}
    \item \textbf{Mailing list archive (January 2001–March 2023)}: Technical discussions and communications were extracted from Gentoo’s official archive website. A total of 127,584 emails and messages were collected. Each entry was annotated with metadata such as the timestamp, sender, recipient and subject line. This dataset was extracted, cleaned and prepared using automated scripts, forming the basis for sentiment analysis that examined developer interactions across different phases of the project.
    
    \item \textbf{Commit history dataset (July 2000–March 2023)}: Commit logs were retrieved directly from Gentoo's GitHub repository using the \textit{`git clone`} command. The dataset was structured into a database containing 5,500,779 commits. Each commit was annotated with key metadata, including committer details (e.g., name and email), timestamps, commit hashes and file paths. These details enabled precise mapping of developer activities to specific categories within the Gentoo ecosystem.
\end{enumerate}

The mailing list archive and commit history were integrated to enable a general analysis of developer sentiment and its relationship with activity patterns. This integration relied on consistent metadata, including email addresses and timestamps, allowing the linkage of sentiment data with technical contributions.

The mailing list data served as the primary source for sentiment analysis, while the commit history dataset was used to quantify the developer turnover, retention and growth rates. Together, these datasets provided a valuable resource for examining sentiment-affected components, enabling longitudinal insights into the evolving dynamics of the Gentoo ecosystem. 

\subsection{Data Preprocessing}
Prior to analysis, both the mailing list and commit datasets underwent extensive preprocessing to ensure accuracy and consistency. 

\paragraph{Mailing list preprocessing} - 
The content of each email from the mailing list dataset was extracted and divided into individual sentences. To remove noise and irrelevant information, we applied the following cleaning steps:
\begin{itemize}
    \item Removed sentences prefixed by the character ‘\textgreater’, which typically denote quoted replies.
    \item Stripped URLs, names or signatures and greeting phrases (e.g., “Kind Regards” and “Best Regards”).
    \item Eliminated sentences containing code syntax, HTML, or XML tags.
\end{itemize}
This process resulted in a clean dataset of 662,731 sentences, each assigned a sentiment score using the SentiStrength-SE tool. These sentiment-annotated sentences were stored in a separate sentiment table to facilitate subsequent analysis.

\paragraph{Integration of mailing list and commit data} - 
To address the research questions (RQ1-RQ2), it was essential to link sentiment data from the mailing list with commit data from GitHub. This required resolving inconsistencies in developer identifiers, as many developers used multiple email addresses across datasets. The following standardization steps were applied:
\begin{itemize}
    \item Unified email addresses by selecting a primary address for each developer, ensuring consistent mappings between sentiment and commit data.
    \item For email addresses lacking explicit names, derived names from the account identifiers within the email addresses.
    \item Standardized developer names across multiple formats, adopting a single canonical format for consistency.
\end{itemize}

After these preprocessing steps, each developer's email and name were consistently represented in both datasets, enabling accurate linkage. Additionally, file path names from the commit dataset were preprocessed to extract \textit{category} names, ensuring compatibility with sentiment-based analysis.

\paragraph{Significance for research questions} - 
These preprocessing steps were critical for linking sentiment-driven developer interactions with commit-based activity metrics, forming the basis for the study’s investigation of turnover, retention, and growth rates and sentiment-affected components.


\subsection{Data Analysis}
\label{sec:_data_analysis}
The analysis phase of this study involved data aggregation and statistical techniques to investigate the relationship between developer sentiment and activity within the Gentoo ecosystem. Specifically, we analyzed both negative and positive sentiments expressed in mailing list communications and their associated activity levels in the commit dataset.

\begin{figure}[ht!]
    \centering
    \includegraphics[width=0.85\linewidth]{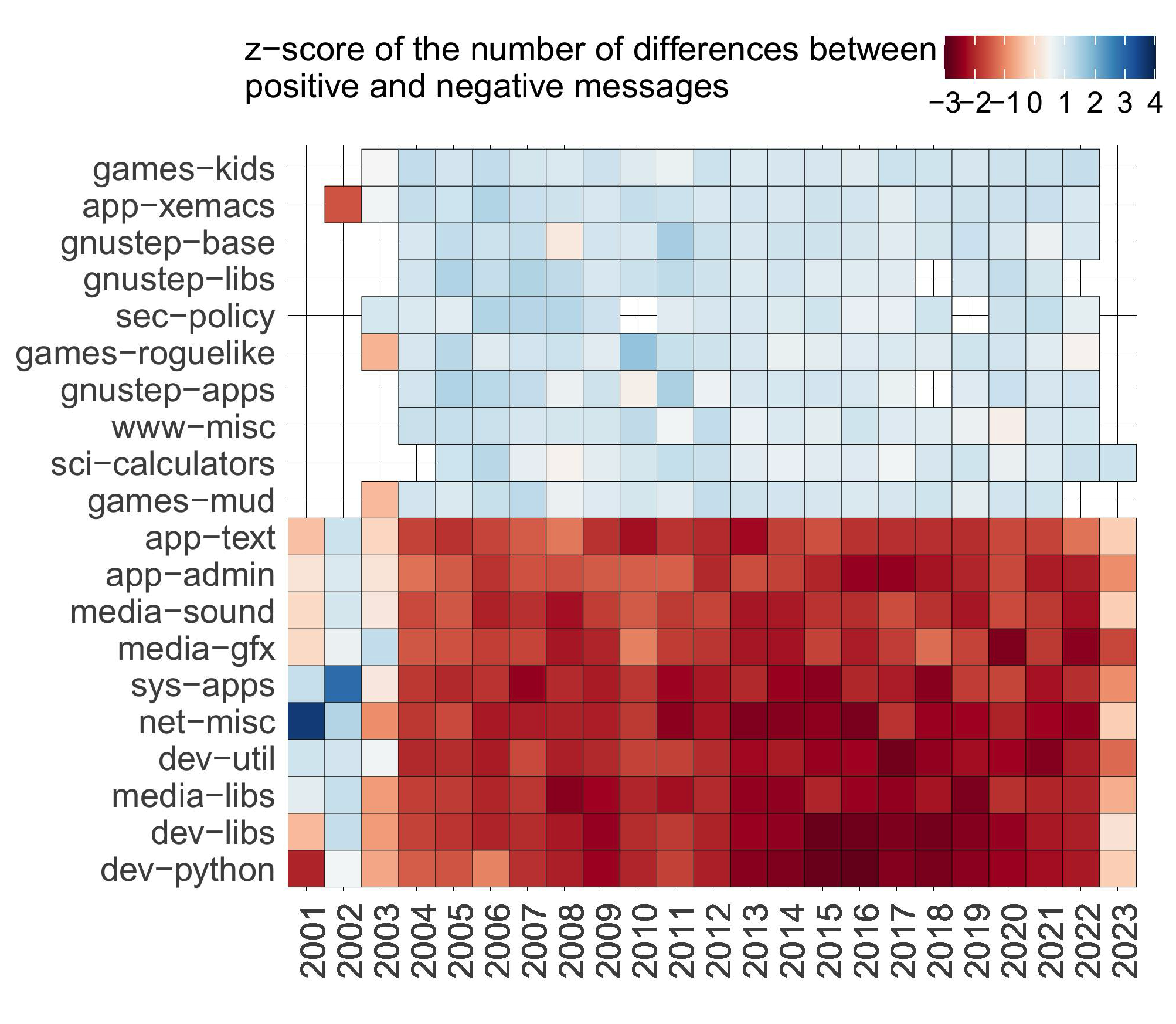}
    \caption{Heatmaps of messages containing negative/positive sentences in the 20 studied categories} 
    \label{fig_RQ2:_fig_20_paths_normalised_yearly} 
\end{figure}

\paragraph{Linking and aggregating sentiment data} -
To address RQ1, we integrated sentiment data with commit activity by aligning records based on timestamps (year, month, day). This integration allowed us to aggregate the number of positive and negative messages for each file path annually. To evaluate sentiment trends over time, we calculated the net sentiment difference (positive minus negative messages) and normalized the results using z-score normalization. Categories were ranked from most negative to the most positive, on a scale of -3 and 4, between 2001 and 2023: the results are visualized in the heatmap plot shown in Figure~\ref{fig_RQ2:_fig_20_paths_normalised_yearly}\footnote{All the raw material and data is provided for replication in this link: https://tinyurl.com/supplement-files}.

\paragraph{Sampling categories for analysis} - 
Given Gentoo's extensive repository of 170 categories as of our analysis started, we selected a representative 10\% sample, resulting in 20 categories for detailed analysis. These included the ten most SN and the ten most SP categories based on sentiment scores. This approach ensured a balanced investigation of extremes in sentiment-affected components while maintaining analytical feasibility.

\paragraph{Calculating workforce metrics} -
We analyzed three key workforce metrics—retention, turnover and growth rates—within the sampled categories, defined above.

\paragraph{Statistical analysis for RQ1} - 
\textbf{(a) Comparison between sentiment groups:} To evaluate differences in retention, turnover, and growth rates between positive and negative categories, we applied the Mann-Whitney U test. Since our data is not normally distributed, we chose this non-parametric test for its robustness in handling non-normal distributions in the data.

\textbf{(b) Temporal analysis within categories:} To investigate variations across distinct time periods (\textit{Early}, for the 2001–2007 interval, \textit{Middle} for 2008–2014 and \textit{Latter} for 2015–2023) we performed pairwise comparisons using the Mann-Whitney U test with the Bonferroni correction. We utilized the Bonferroni correction as our work conducted multiple pairwise comparisons, in which the probability of yielding one or more false positives (Type I error) is higher. Time periods were defined in approximately seven-year intervals, with data from 2023 included up to March. To visualize these findings, boxplots with jitter were generated to illustrate differences and distributions across periods.

\paragraph{Correlation analysis for RQ2} -
To explore the potential relationship between workforce metrics and changes in sentiment-affected components, we employed a correlation matrix of Pearson method with retention, turnover and growth rates and the number of file modifications as the variables measured. We utilized the correlation matrix approach as our values of variables measured are numerical. In addition, we intend to quantify both the strength and direction of the relationship between two variables. 
The analysis included:
\begin{itemize}
    \item \textbf{Data transformation:} As we calculate the number of modifications M as defined in Section~\ref{sec:_definitions} and the distribution of the independent variables measured in the formula are not normal, we normalized all values of N, C\_retained, G and C\_new by transforming their values function logarithmic or log(n). 
    \item \textbf{Software tools:} All statistical analyses were conducted using R version 4.2.2.
\end{itemize}


These methodologies provided a framework for exploring the dynamics of sentiment and workforce metrics within the Gentoo ecosystem, addressing the research questions and yielding actionable insights for managing open-source projects.

\section{Results}
\label{sec:_results}
\vspace{-0.5em}
\subsection{RQ1 (a) - Turnover (TR), retention (RR), growth rates (GR)}
\vspace{0.5em}

Significant differences in workforce dynamics (turnover, retention, and growth rates) were observed between the SN and SP categories. 

First, turnover rates were analyzed using the Mann-Whitney U test: results revealed that turnover rates differ significantly between SN and SP categories, with a p-value < $2.2 \times 10^{-16}$ after applying the Bonferroni correction. Consequently, we reject the null hypothesis $H_{RQ1A_1,0}$, confirming a statistically significant difference in turnover rates between the two groups.

Similarly, retention rates were assessed using the same statistical method: the p-value, again < $2.2 \times 10^{-16}$ after Bonferroni correction, led to rejecting the null hypothesis $H_{RQ1A_2,0}$. This indicates that also retention rates are significantly different across SN and SP categories.

Finally, growth rates were analyzed through the Mann-Whitney U test: a p-value < $2.2 \times 10^{-16}$ after Bonferroni correction enabled the rejection of the null hypothesis $H_{RQ1A_3,0}$, indicating a furtherstatistically significant difference in growth rates between the SN and SP categories.

Additionally, pairwise comparisons were conducted for each workforce rate across the three defined time phases, further substantiating these results.

\vspace{0.3cm}
\fbox{
\parbox{0.9\columnwidth}{
\textbf{Summary:} 
Workforce dynamics (turnover, retention, and growth rates) differ significantly between SN and SP categories. Pairwise comparisons across time phases further substantiate these findings, highlighting distinct patterns in workforce behavior.
}
}

\subsection{RQ1 (b) - Dynamic analysis of turnover, retention, growth rates.}
\vspace{0.5em}

\begin{figure*}[ht!]
    \centering
    \includegraphics[scale=0.6]{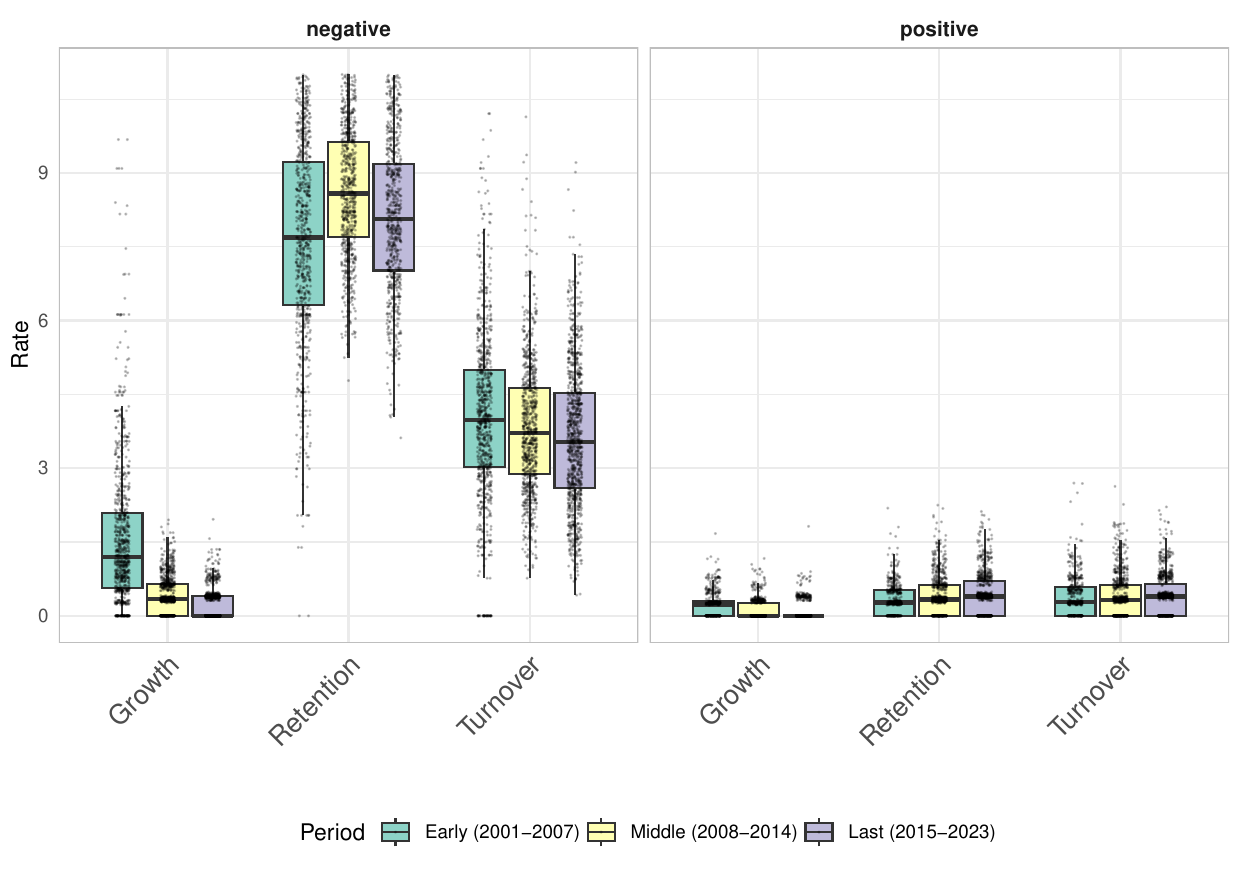}
    \caption{Boxplots of Turnover, Retention and Growth Rate in three different time periods by sentiment-affected groups}
    \label{fig:_boxplots_jitter_two_groups}
\end{figure*}

Figure~\ref{fig:_boxplots_jitter_two_groups} illustrates the distribution of growth, retention, and turnover rates across the SN and SP categories: these groups were further analyzed over three distinct time periods: early phase (2001--2007), middle phase (2008--2014), and last phase (2015--2023). The figure reveals significant differences in workforce metrics between the two groups, notably the median values for all metrics in the SN group are approximately double those observed in the SP group.

To support the evidence around these differences, we conducted Mann-Whitney U tests with Bonferroni corrections. All null hypotheses were rejected, with adjusted p-values below the threshold alpha ($\alpha$/9, 0.00556). Specifically, we rejected $H_{RQ1B_1,0}$, $H_{RQ1B_2,0}$, and $H_{RQ1B_3,0}$, indicating significant disparities across all workforce metrics and time intervals.

Examining the retention rate within the SN group, its median consistently exceeds 7\%, surpassing both growth and turnover rates, which generally range between 1\% and 5\%. Over time, both growth and turnover rates show a decreasing trend. In contrast, SP categories exhibit considerably lower median values across all metrics, remaining below 1\%. Here, retention and turnover rates are marginally higher than growth rates.

Table~\ref{tab_pairwise_comparisons} provides the results of pairwise comparisons (denoted as $\delta_1$, $\delta_2$, and $\delta_3$) for RR, TR, and GR within SN and SP categories. The findings highlight substantial intra-group differences across phases, with over 50\% (48 out of 90) of tests revealing significant p-values ($p < 0.05/90$).

\begin{table}[!ht]
\centering
\caption{Pairwise comparisons with U Mann-Whitney test and Bonferroni Correction within SN and SP categories}
\vspace{-1em}

\label{tab_pairwise_comparisons}
\small
\begin{threeparttable}
    \begin{tabular}{|p{2cm}|c|c|c|c|c|c|c|c|c|}
    \hline
     {} & \multicolumn{3}{c|}{RR} & \multicolumn{3}{c|}{TR} & \multicolumn{3}{c|}{GR} \\
     \cline{2-10} 
         {} & $\delta1$ & $\delta2$ & $\delta3$ & $\delta1$ & $\delta2$ & $\delta3$ & $\delta1$ & $\delta2$ & $\delta3$ \\
    \hline
        \multicolumn{2}{|c}{\textbf{SN categories}} & \multicolumn{8}{c|}{}\\    
    \hline
        dev-python & \cellcolor{red!25}\textbf{Y} & \cellcolor{red!25}\textbf{Y} & \cellcolor{red!25}\textbf{Y} & N & N & \cellcolor{red!25}\textbf{Y} & \cellcolor{red!25}\textbf{Y} & \cellcolor{red!25}\textbf{Y} & \cellcolor{red!25}\textbf{Y} \\
    \hline
        dev-libs & \cellcolor{red!25}\textbf{Y} & \cellcolor{red!25}\textbf{Y} & \cellcolor{red!25}\textbf{Y} & N & N & N & \cellcolor{red!25}\textbf{Y} & \cellcolor{red!25}\textbf{Y} & \cellcolor{red!25}\textbf{Y} \\
    \hline
        media-libs & N & \cellcolor{red!25}\textbf{Y} & N & N & \cellcolor{red!25}\textbf{Y} & \cellcolor{red!25}\textbf{Y} & \cellcolor{red!25}\textbf{Y} & N & \cellcolor{red!25}\textbf{Y} \\
    \hline
        dev-util & \cellcolor{red!25}\textbf{Y} & \cellcolor{red!25}\textbf{Y} & \cellcolor{red!25}\textbf{Y} & N & N & N & \cellcolor{red!25}\textbf{Y} & N & \cellcolor{red!25}\textbf{Y} \\
    \hline
        net-misc & N & N & N & N & N & \cellcolor{red!25}\textbf{Y} & \cellcolor{red!25}\textbf{Y} & \cellcolor{red!25}\textbf{Y} & \cellcolor{red!25}\textbf{Y} \\
    \hline
        sys-apps & \cellcolor{red!25}\textbf{Y} & N & \cellcolor{red!25}\textbf{Y} & N & N & \cellcolor{red!25}\textbf{Y} & \cellcolor{red!25}\textbf{Y} & N & \cellcolor{red!25}\textbf{Y} \\
    \hline
        media-gfx & N & N & N & N & N & \cellcolor{red!25}\textbf{Y} & \cellcolor{red!25}\textbf{Y} & N & \cellcolor{red!25}\textbf{Y} \\
    \hline
        media-sound & N & \cellcolor{red!25}\textbf{Y} & \cellcolor{red!25}\textbf{Y} & N & \cellcolor{red!25}\textbf{Y} & \cellcolor{red!25}\textbf{Y} & \cellcolor{red!25}\textbf{Y} & N & \cellcolor{red!25}\textbf{Y} \\
    \hline
        app-admin & \cellcolor{red!25}\textbf{Y} & N & \cellcolor{red!25}\textbf{Y} & N & N & N & \cellcolor{red!25}\textbf{Y} & N & \cellcolor{red!25}\textbf{Y} \\
    \hline
        app-text & N & N & N & N & \cellcolor{red!25}\textbf{Y} & N & \cellcolor{red!25}\textbf{Y} & \cellcolor{red!25}\textbf{Y} & N\\
    \hline
        \multicolumn{2}{|c}{\textbf{SP categories}} & \multicolumn{8}{c|}{}\\    
    \hline
        games-kids & N & N & N & N & N & N & N & N & \cellcolor{red!25}\textbf{Y} \\
    \hline
        app-xemacs & N & N & \cellcolor{red!25}\textbf{Y} & N & N & N & N & N & \cellcolor{red!25}\textbf{Y} \\
    \hline
        gnustep-libs & N & N & N & N & N & N & N & N & N \\
    \hline
        gnustep-base & N & N & N & N & N & N & \cellcolor{red!25}\textbf{Y} & N & \cellcolor{red!25}\textbf{Y} \\
    \hline
        sec-policy & N & N & N & N & N & N & N & N & N \\
    \hline
        games-roguelike & N & N & N & N & N & N & N & N & N \\
    \hline
        gnustep-apps & \cellcolor{red!25}\textbf{Y} & N & N & N & N & N & N & N & \cellcolor{red!25}\textbf{Y} \\
    \hline
        www-misc & N & N & N & N & N & N & N & N & \cellcolor{red!25}\textbf{Y} \\
    \hline
        sci-calculators & N & N & N & N & N & N & \cellcolor{red!25}\textbf{Y} & N & \cellcolor{red!25}\textbf{Y} \\
    \hline
        games-mud & N & \cellcolor{red!25}\textbf{Y} & \cellcolor{red!25}\textbf{Y} & N & N & N & N & N & N \\
    \hline
    \end{tabular}
    \begin{tablenotes}[flushleft]
        \small
        \item RR, TR, GR, $\delta1$,  $\delta2$ and  $\delta3$: please see the definitions in Section ~\ref{sec:_definitions}.
        \item Y means p-value < 0.05/90; N means p-value > 0.05/90 after the Bonf. correction.
    \end{tablenotes}
\end{threeparttable}
\end{table}

Detailed analysis shows that the most SN category, \textit{dev-python}, has seven out of nine comparisons with significant p-values. RR and GR exhibit notable differences across all intervals ($\delta_1$, $\delta_2$, and $\delta_3$). Turnover rates, while stable, also display significant variation in certain intervals, such as $\delta_3$. Similarly, \textit{dev-libs} and other SN categories, including \textit{media-libs}, \textit{dev-util}, \textit{sys-apps}, and \textit{media-sound}, exhibit significant differences in various metrics, reinforcing the heterogeneity within this group.

On the contrary, SP categories demonstrate limited variation. Out of 90 tests, only twelve significant p-values were observed. For instance, \textit{gnustep-base} and \textit{sci-calculators} showed significance in $\delta_1$ and $\delta_3$ for growth rates, while \textit{games-mud} displayed a significant result in $\delta_2$  and $\delta_3$ for retention rate. Furthermore, a single significant result spread in each metric occurred in the rest of the categories. These results suggest greater stability but reduced dynamism within SP categories.

The findings underscore the noticeable dynamics in SN categories, particularly regarding retention and growth rates, which vary across phases and categories. In contrast, SP categories exhibit minimal modifications, highlighting the need for tailored management strategies to address the contrasting stability and innovation challenges of each group.
\vspace{0.2cm}

\fbox{
\parbox{0.9\columnwidth}{
\textbf{Summary:} 
SN categories show higher and more dynamic workforce metrics (retention, growth, turnover) compared to SP categories, which remain stable but low. Key variations in SN groups suggest adaptability challenges, while SP groups prioritize stability. Tailored strategies are needed for balance and sustainability.}
}

\subsection{RQ2 - Correlation between number of modifications (M), TR, RR and GR}
\vspace{0.5em}

\begin{figure}[ht!]
    \includegraphics[width=0.95\columnwidth]{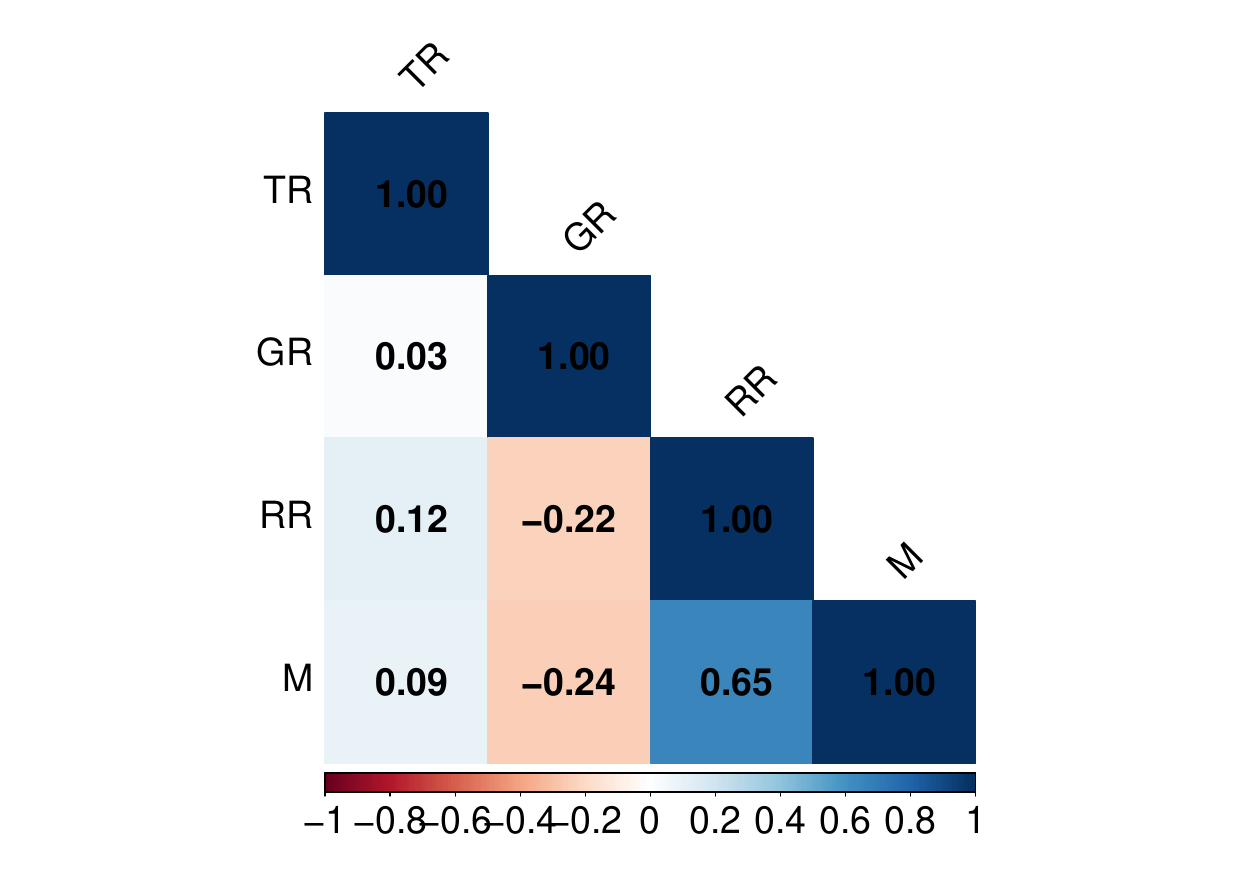}
    \caption{Correlation matrix between M and TR, GR, and RR in SN categories.}
    \label{fig:_correlation_matrix}
\end{figure}

\begin{figure}[ht!]
    \includegraphics[width=0.95\columnwidth]{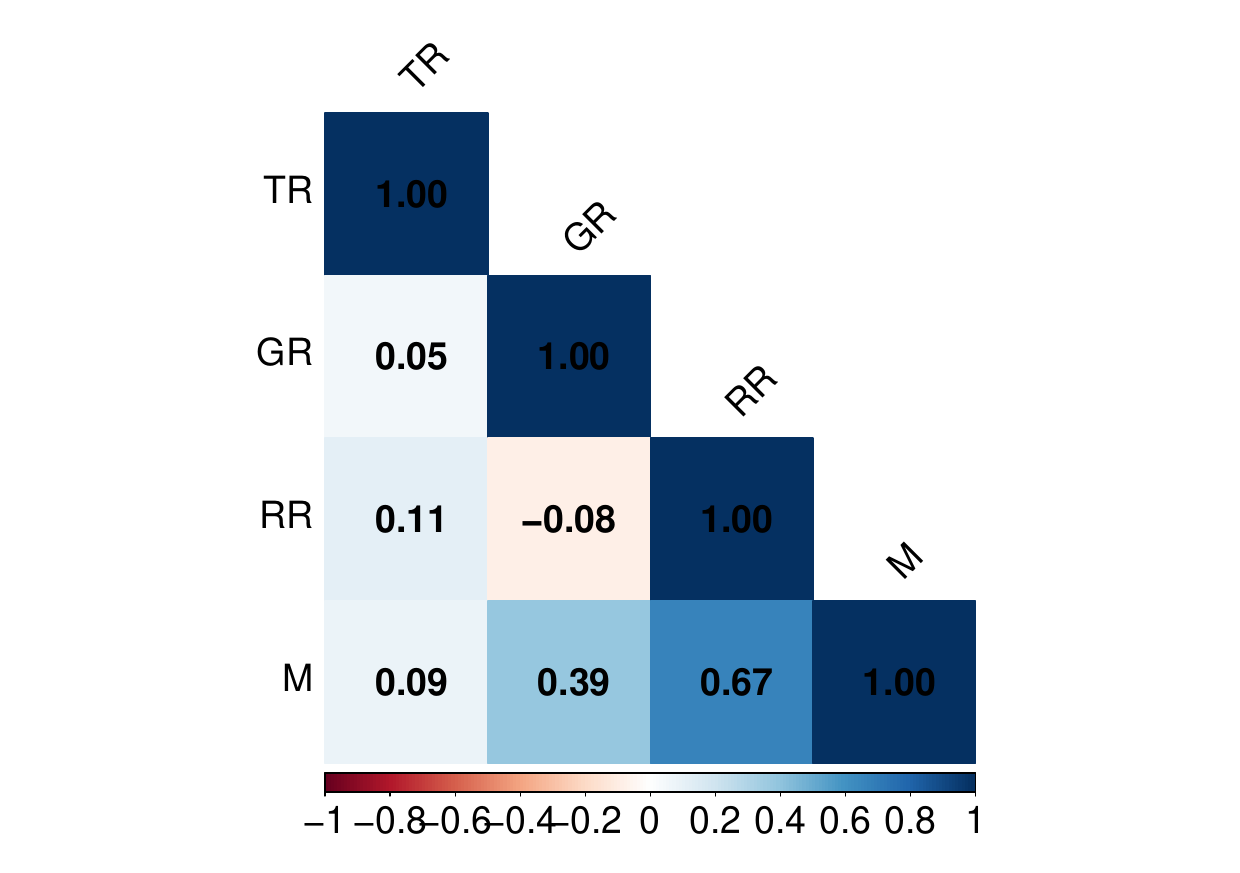}
    \caption{Correlation matrix between M and TR, GR, and RR in SP categories.}
    \label{fig:_correlation_matrix_SP}
\end{figure}

All correlation metrics between $M$, $TR$, $RR$, and $GR$ (see Figures \ref{fig:_correlation_matrix} and \ref{fig:_correlation_matrix_SP}) were quite similar between the SN and SP categories, except for the relationships highlighted below:

\paragraph{Relationship between $M$ and $GR$} - In the SN components, the negative correlation ($r = -0.24$) indicates a more stable environment, where growth is aligned to a reduced modification activity. This is likely due to improved code quality and fewer major changes, which improves long-term sustainability.
In SP components, a positive correlation ($r = 0.39$) suggests that growth drives an increased modification activity. This could reflect a reactive response to challenges, or workload distribution. However, this dynamic could signal instability, since high growth often is related to high turnover.

\paragraph{Relationship between $RR$ and $GR$} - In SN components, an $r = -0.22$ suggests that growth undermines retention, reflecting challenges in integrating new contributors. In contrast, the $r = -0.08$ of SP components indicates a stable and collaborative workforce, where growth has minimal impact on retention.

\vspace{-10pt}

M has a strong positive correlation with RR in both SN and SP groups, in which r = 0.65 and r = 0.67, respectively. These correlations may indicate that the more developers continued working, the more changes were made regardless of the sentiment-affected components where they worked. Conversely, M has no association with TR in both SN and SP groups, with r = 0.09 each. 
\vspace{0.3cm}

\fbox{
\parbox{0.9\columnwidth}{
\textbf{Summary:} 
In SN components, growth negatively correlates with modifications; in SP components, this correlation is positive. Growth undermines retention in SN categories, but has minimal impact in SP categories. Retention strongly correlates with modifications in SN and SP categories, while turnover shows no association with modifications in either group.
}
}

\section{Discussion and Implications}
\label{sec:_discussion_and_implications}
Our analysis revealed significant differences in workforce metrics between SN and SP categories, highlighting the impact of sentiment on workforce dynamics. SN categories exhibited higher turnover (TR), retention (RR), and growth rates (GR) compared to SP categories, suggesting distinct patterns influenced by sentiment.

The increased TR in SN categories points to workforce instability, potentially caused by dissatisfaction or poor communication. This confirms Chaves et al.'s work~\cite{chaves2022autonomy} that one factor contributing to turnover was the communication contributor's opinion. Furthermore, the higher turnover rate observed in SN categories compared to SP categories is supported by previous research, which found a positive relationship between negative affect and withdrawal behavior~\cite{pelled1999down}.

Despite this, high RRs indicate the presence of a core group of dedicated developers who sustain these components, likely due to their specialized roles or the critical importance of their contributions. This specialized core supports continuity as components stabilize over time. The social aspects (e.g. helping others and teamwork), learning, and intellectual stimulation may contribute to the high RR~\cite{Zhou2012What}. Our finding aligns with the work of Balthrop and Jung; the fulfilled mutual interests between negative employees and employers lead to employee embeddedness~\cite{balthrop2024shared}.

Higher GR in SN categories reflects active recruitment efforts to address turnover and maintain productivity~\cite{Daniel2009Using}. However, this strategy may present challenges in onboarding and retaining new contributors, underscoring the need for tailored approaches to workforce integration in such environments.

Most categories displayed pronounced dynamics, specifically RR and GR, particularly during later phases. These variations likely stem from factors such as team size, complexity, or community culture. In contrast, categories with minimal variation in these metrics represent more stable or less dynamic contexts.

High RRs emerged as critical for stability in both sentiment groups, enabling developers to focus on meaningful contributions. In SN categories, addressing high TR while bolstering RR could reduce disruptions and improve performance. In SP categories, strong retention underpins sustained development activity, suggesting these environments are less vulnerable to instability from turnover. Regardless of the sentiments involved during the development, our findings align with the prior study that the opportunities to learn new skills are the main reason for the motivational process for developers to thrive at work~\cite{trinkenreich2024predicting}.

Interestingly, the low correlation between development activity and TR in both sentiment groups indicates that turnover may be driven by external factors beyond immediate project demands. Future research should examine the role of team dynamics, project management practices, and organizational influences on workforce stability. Additionally, the relationship between RR and development activity highlights RR as a key driver of engagement and continuity, though broader contextual factors may also play a role.

\subsection{Implications for Practitioners}
The observed relationship between retention rates and the number of file changes emphasizes the crucial role of retaining developers to sustain project \textit{momentum} and stability. Practitioners should prioritize fostering a supportive and engaging work environment to enhance job satisfaction and work-life balance, thereby mitigating attrition risks. Recognition programs for developers who remain committed to long-term projects, especially in SN components, can reinforce their contributions and sustain morale in challenging environments.

Turnover and retention metrics are key tools for identifying issues at both component and project levels. Implementing systems to track workforce dynamics alongside productivity metrics allows for the proactive detection of instability. Real-time monitoring can highlight areas with declining retention or high turnover, enabling targeted interventions to stabilize teams and maintain development activity.

For SN components, which often show high modification activity but face workforce instability, management should foster a collaborative environment. Clear communication, conflict resolution, and mentoring can address turnover issues. Retention strategies should focus on building a core group of contributors to ensure continuity and prevent stagnation.

For SP components, while benefiting from inherent stability, are at risk of stagnation due to a lack of dynamic growth. Managers should emphasize fostering innovation by encouraging experimentation, setting clear growth targets, and ensuring consistent contributor engagement. This balance can help maintain a dynamic, stable development environment, allowing SP components to thrive without sacrificing long-term sustainability.

Overall, practitioners are encouraged to tailor their management strategies to the unique dynamics of sentiment-driven components, leveraging these insights to improve workforce stability, project sustainability and innovation within open-source ecosystems.
\vspace{-0.5em}

\subsection{Implications for Researchers}
The findings of this study underscore the importance of exploring the causal mechanisms behind the observed correlations between sentiment-driven dynamics and workforce behavior. While this research highlights \textit{correlations}, future studies should delve deeper into why growth rate negatively impacts development activity in SN components (e.g., \textit{causation}) or why the turnover rate is predominantly exhibited in these components. Identifying the challenges faced by contributors in these environments could provide insights for enhancing stability, attracting developers, and reducing turnover. Researchers might explore how retention in SP components boosts development activity, offering new perspectives on productivity and sustainability.

Improving predictive models offers a valuable opportunity for researchers. Future studies should consider various factors, such as developer experience, geographic distribution, project involvement, and software characteristics like complexity and bug report frequency. By expanding the analysis, researchers can create more effective workforce dynamics models.

Cross-ecosystem studies are vital for validating and generalizing findings. By including open-source projects with diverse governance structures, cultures, and technologies, we can better understand how sentiment and workforce dynamics interact. Additionally, examining temporal trends can reveal long-term patterns in workforce behavior, enhancing the literature on developer sentiment and project sustainability.

Finally, researchers are encouraged to utilize this study as a foundation for further inquiries into the interplay between sentiment, workforce stability, and development outcomes in open-source ecosystems. By building on these insights, the research community can contribute to advancing the sustainability and resilience of open-source projects globally.
\vspace{-0.5em}

\section{Threats to validity}
\label{sec:_threats_to_validity}
\paragraph{\textbf{Internal validity}} - 
A primary threat to internal validity arises from the reliance on SentiStrength-SE for sentiment analysis. While effective for many sentiment classification tasks, this tool may misclassify nuanced expressions, such as sarcasm or idiomatic phrases, potentially introducing bias. Incorporating additional sentiment analysis methods could mitigate this limitation.

Inaccuracies in linking mailing list data with commit data present another challenge. Developers often use multiple email addresses or aliases, which may lead to incomplete or erroneous mappings. Although standardization techniques were applied, residual inconsistencies may affect the accuracy of workforce metrics such as turnover and retention rates.

Finally, while the study establishes correlations between workforce dynamics and software stability, causality cannot be definitively inferred. Co-founding factors, such as component size, team distribution, or technical complexity, might influence the observed relationships, requiring caution when interpreting results.

\paragraph{\textbf{External validity}} - 
The study's findings are derived from the Gentoo open-source ecosystem, which has unique governance structures and contributor dynamics. As a result, the generalizability of these insights to other ecosystems with different characteristics may be limited. Expanding the scope of future research to encompass other open-source projects would help validate these findings.

The extended 23-year observation period introduces potential biases: the evolution of software development practices, tools and community behaviors over this time may not be uniformly captured, which could affect the temporal generalizability of the results.

Variations in cultural and linguistic expressions within the Gentoo ecosystem may influence the manifestation of sentiment-driven dynamics. This limitation suggests a need for further studies in more culturally diverse settings to better understand these effects.

\paragraph{\textbf{Construct validity}} - 
The operationalization of sentiment groups and workforce metrics poses potential threats to construct validity. The binary classification of components into SN or SP categories may oversimplify the complex emotional landscapes of developer interactions. Future work could explore a continuum-based sentiment categorization to better reflect nuanced emotional states.

The use of retention, turnover and growth rates as proxies for workforce dynamics is another limitation. These metrics, while useful, may not fully account for other critical aspects such as developer productivity, workload balance, or satisfaction. Including additional indicators could enhance the robustness of the findings.

Lastly, the focus on the top 10 positive and negative sentiment categories excludes neutral or mixed sentiment components, which could provide valuable comparative insights. Expanding the analysis to include these categories would offer a more holistic view of sentiment-driven dynamics within the ecosystem.
\vspace{-0.5em}

\section{Conclusion and Future Work}
\label{sec:_conclusions}
This study investigated the interplay between developer sentiment and workforce dynamics within the Gentoo open-source ecosystem over a 23-year period: by analyzing turnover, retention and growth rates across sentiment-negative (SN) and sentiment-positive (SP) components, our study revealed interesting insights into the impact of sentiment on project stability and evolution. SN components were found to exhibit higher retention rates but lower turnover and growth, suggesting stability at the cost of innovation; on the other hand, SP components demonstrated higher turnover rates than growth rates at the base of a dynamic yet potentially less stable environment.

These findings underscore the necessity of tailoring management strategies to sentiment dynamics: for SN components, strategies focused on increasing adaptability and innovation could prevent risks of stagnation. On the other hand, SP components could benefit from enhanced mechanisms to improve long-term contributor retention, in this way addressing sustainability concerns.

Future research should extend this analysis to other open-source ecosystems to validate the generalizability of these findings across diverse governance structures and development practices. Additionally, incorporating more granular metrics, such as `developer productivity' and `satisfaction' (alongside sentiment and workforce dynamics) could produce a more comprehensive understanding of the factors driving stability and growth. Investigating the influence of cultural, organizational, and technological changes on sentiment and workforce behaviors over time also represents a promising avenue for future exploration. 

\bibliographystyle{abbrvnat}

\end{document}